\documentclass[12pt,a4paper]{article}
\usepackage[T1]{fontenc}        
\usepackage{amsmath,amssymb,dsfont}    
\usepackage[dvips]{epsfig}
\title{
  {\vspace{-2cm} \normalsize
     \epsfig{figure=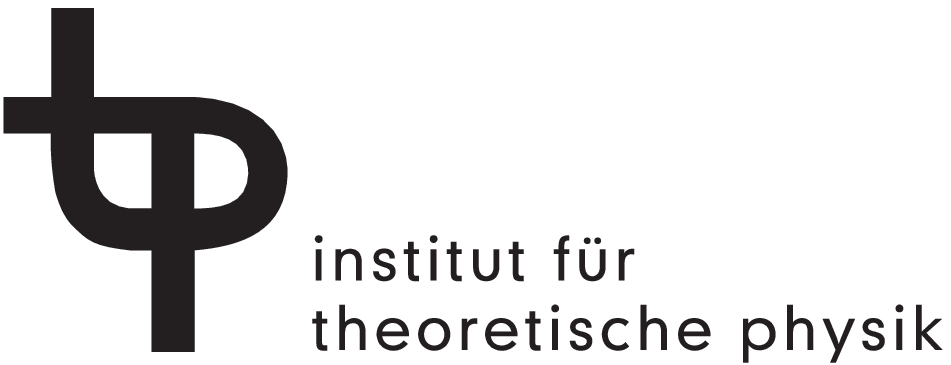,width=80mm}
     \hfill\parbox[b][30mm][t]{35mm}{MS-TP-06-01 \\
                                     hep-lat/0603019}  }\\[30mm]
  Twisted mass lattice QCD\\
  with non-degenerate quark masses
  }
\author{Gernot M\"unster and Tobias Sudmann\\
        Institut f\"ur Theoretische Physik,
        Universit\"at M\"unster\\
        Wilhelm-Klemm-Str.~9, D-48149 M\"unster, Germany\\
        e-mail: munsteg@uni-muenster.de, tobisud@uni-muenster.de}
\date{March 22, 2006 (revised April 14, 2006)}
\newcommand{\I}{\ensuremath{\mathrm{i}\,}}
\newcommand{\E}{\ensuremath{\mathrm{e}\,}}
\newcommand{\Tr}[1]{\ensuremath{\operatorname{Tr}\left(#1\right)}}
\newcommand{\dmu}{\ensuremath{\partial_\mu}}
\begin{document}
\maketitle
\begin{abstract}
Quantum Chromodynamics on a lattice with Wilson fermions and a chirally
twisted mass term is considered in the framework of chiral perturbation
theory.  For two and three numbers of quark flavours, respectively,
with non-degenerate quark masses the pseudoscalar meson masses and
decay constants are calculated in next-to-leading order including
lattice effects quadratic in the lattice spacing $a$.
\end{abstract}
%
When quantum field theories are simulated numerically, the introduction
of a finite space-time lattice brings with it two standard problems. In
order to obtain results relevant to the physics in the infinite
continuum the effects of the finite lattice spacing $a$ and of the
finite volume of the system have to be controlled. The resulting
demands on the lattice are limited by the available computer resources.
Simulations of Quantum Chromodynamics are facing a third difficulty.
The common algorithms slow down severely when the masses of the light
quarks are decreased towards their physical values.

Recent investigations \cite{TWIST,DBW,DBW2} suggest that lattice QCD
with Wilson fermions and a chirally twisted quark mass term
\cite{TM,Frezzotti} offers an attractive framework for numerical
simulations with small up- and down-quark masses.  In particular, for
the case of full twist lattice artifacts of order $a$ are removed
automatically \cite{Frezzotti-Rossi}.  For a review of twisted mass
lattice QCD see \cite{Shindler}.

An important tool for QCD is chiral perturbation theory
\cite{Weinberg,GL1,GL2}. For sufficiently small quark masses it
describes the dependence of mesonic quantities like masses and decay
constants on the quark masses in terms of expansions in powers of quark
masses (modified by logarithms). These expansions contain low energy
parameters, the \textit{Gasser-Leutwyler coefficients}, which have to
be determined by methods outside chiral perturbation theory.

Chiral perturbation theory is particularly suited for lattice QCD. In
its range of applicability, it can be used to extrapolate the results
of Monte Carlo simulations into the region of small quark masses. On
the other hand, numerical simulations of lattice QCD can provide
information about the Gasser-Leutwyler coefficients, see \cite{Wittig}
for a review.

The adaption of chiral perturbation theory to lattice QCD
\cite{ShaSi,LeeSha,Rupak-Shoresh,Aoki,BRS} brings the lattice spacing
$a$ in as a second expansion parameter.  In the effective chiral
Lagrangian lattice artifacts are represented by additional terms
proportional to powers of $a$ \cite{Rupak-Shoresh}.  Physical quantities
appear in double expansions in quark masses and in $a$.

Chiral perturbation theory has been extended to the case of twisted
mass lattice QCD in \cite{MS,MSS}.  The resulting expansions contain
the twist angle as an additional parameter.  Results for various
quantities have been calculated in next-to-leading order of chiral
perturbation theory including lattice effects of order $a^2$
\cite{MS,MSS,Muenster,Sco,SHARPE-WU1,AoBa,SHARPE-WU2,SHARPE2}.

In most of the work the case of degenerate quark masses is considered.
Current numerical work on twisted mass lattice QCD with non-degenerate
quark masses \cite{tmLQCD2} motivates to implement this situation in
chiral perturbation theory.  This is the purpose of the work presented
here.

We consider lattice QCD with $N_f =2$ or $3$ quark flavours and a
twisted mass term. In the two-flavour case we consider split quark
masses ($m_u \neq m_d$). In the case of three flavours we restrict
ourselves in this letter to equal up- and down-quark masses and a
heavier strange quark mass ($m_u = m_d \neq m_s$). The masses and decay
constants of the pseudoscalar mesons are calculated in next-to-leading
(one-loop) order, including lattice terms of order $a^2$.

Chiral perturbation theory for pseudoscalar meson masses in untwisted
lattice QCD with $N_f = 3$ flavours of Wilson quarks has been
considered in \cite{ABTI}.  These authors use a power counting scheme
different from the one of \cite{Rupak-Shoresh,BRS,MS,MSS} and a
different parameterization of the action, which hampers a comparison of
our calculations.  The results which could be compared in the limit of
vanishing twist angle do agree.  The case of twisted mass lattice QCD
with 2+2 flavours has been studied recently by means of chiral
perturbation theory in \cite{ALWW}. The implementation of mass
non-degenerate quarks in twisted mass chiral perturbation theory for
two flavours has been considered by Walker-Loud and Wu \cite{WLW}. They
extend the formalism to the baryon sector and calculate nucleon and
delta masses.

In lattice QCD a chiral twist of the quark mass term is generally
introduced by transforming the quark mass matrix $M$ into
\begin{equation}
M(\omega) =
\E^{\I \omega_b T_b \gamma_5}\, M\, \E^{\I \omega_b T_b \gamma_5}\,,
\end{equation}
where $T_b$ are the generators of SU($N_f$) and $\omega$ denotes the
set of group parameters $\omega_b$. In contrast to the situation in the
continuum, on the lattice the chiral transformation of the mass matrix
has a measurable effect on observables since in lattice QCD with Wilson
fermions chiral symmetry is broken by lattice artifacts.

In chiral perturbation theory the low energy regime of QCD is described
by an effective Lagrangian for the pseudo-Goldstone bosons of
spontaneously broken chiral symmetry. For $N_f=3$ they form the
pseudoscalar meson octet including the pions, kaons and $\eta$.

The transformation properties of the Goldstone bosons in QCD lead to
collect the dynamical variables in the SU($N_f$) matrix valued field
\begin{equation}
U(x) = \exp \left( 2\frac{\I}{F_0} T_b\,\phi_b(x) \right),
\end{equation}
where $\phi_b$ are the components of the meson fields and $F_0$ is a 
low energy parameter.  
Treating quark masses
and lattice effects perturbatively in a spurion analysis with respect to
the required symmetry properties, the leading order effective chiral 
Lagrangian for
lattice QCD is given by~\cite{ShaSi,Rupak-Shoresh}
\begin{equation}
{\cal L}_\mathrm{LO} \equiv {\cal L}_2
= \frac{F_0^2}{4} \Tr{\dmu U \, \dmu U^\dagger }
- \frac{F_0^2}{4} \Tr{\chi U^\dagger + U \chi^\dagger}
- \frac{F_0^2}{4} \Tr{\rho \, U^\dagger + U \rho^\dagger},
\end{equation}
consisting of the chiral symmetric kinetic term and two chiral symmetry 
breaking terms. The first of them contains
the quark mass term, given by
\begin{equation}
\chi = 2B_0 M ,
\end{equation}
and in the second one
lattice artifacts are represented by
\begin{equation}
\rho = 2 W_0 a \mathds{1} ,
\end{equation}
where $B_0$ and $W_0$ are
additional parameters.

Note that we use the power-counting scheme of
\cite{Rupak-Shoresh,BRS}, accounting for terms of order
$\{m,a\}$ in leading order and $\{m^2,ma,a^2\}$ in next-to-leading
order.  In order to define an applicable scheme we always assume the
lattice effects to be smaller:  $W_0 a\ll B_0 m_q$.

Introducing a chiral twist, the mass matrix in the effective Lagrangian
becomes
\begin{equation}
\chi(\omega) =
2B_0\, \E^{-\I \omega_b T_b}\, M\, \E^{-\I \omega_b T_b} \,,
\end{equation}
in the so-called \textit{twisted basis}.  The $\omega_b$-dependence of
the mass term can be removed by changing to the \textit{physical
basis}, i.e.\ by a chiral transformation of the configuration matrix
\begin{equation}
U \longrightarrow U' = \E^{\I \omega_b T_b}\, U\, \E^{\I \omega_b T_b}
\end{equation}
and a corresponding chiral twist of the lattice term according to
\begin{equation}
\rho \longrightarrow \rho(\omega)
= \E^{\I \omega_b T_b}\, \rho\, \E^{\I \omega_b T_b} \,.
\end{equation}
Omitting the prime on the meson field and suppressing the explicit
$\omega_b$-dependence of the lattice term, the effective Lagrangian
looks like the one of untwisted lattice QCD. However, the introduction
of a chirally twisted mass term leads to a shift of the minimum
$\tilde{\phi}\neq 0$ of the effective Lagrangian of order $a$. Chiral
perturbation theory, which is based on an expansion around this
minimum, has to be adapted accordingly. We treat the nontrivial minimum
perturbatively according to the applied scheme.  Denoting the minimum
\begin{equation}
U_0 = \exp \left( 2\frac{\I}{F_0} T_b \, \tilde\phi_b \right),
\end{equation}
a substitution
\begin{equation}
U \longrightarrow U_0^{1/2}\, U\, U_0^{1/2}
\end{equation}
is performed before expanding the Lagrangian. The tree-level and
one-loop contributions of the resulting ${\cal L}_2$ have the same form
as those of continuum chiral perturbation theory, but with different
leading order meson masses which include lattice artifacts, see below.

The next-to-leading order part of the effective Lagrangian in the
present power counting scheme is given by \cite{BRS,MS}
\begin{align}
{\cal L}_4
= & - L_1 \left[\Tr{\dmu U \dmu U^\dagger}\right]^2
- L_2 \left[\Tr{\dmu U \partial_\nu U^\dagger}\right]^2
- L_3 \Tr{[\dmu U \dmu U^\dagger]^2} \notag\\
& + L_4 \Tr{\dmu U \dmu U^\dagger} \Tr{\chi^\dagger U + U^\dagger\chi}
+ W_4 \Tr{\dmu U \dmu U^\dagger} \Tr{\rho^\dagger U + U^\dagger \rho} 
  \notag\\
& + L_5 \Tr{\dmu U \dmu U^\dagger (U \chi^\dagger + \chi U^\dagger)}
+ W_5 \Tr{\dmu U \dmu U^\dagger (U \rho^\dagger + \rho \, U^\dagger)} 
  \notag\\
& - L_6 \left[\Tr{\chi^\dagger U + U^\dagger\chi}\right]^2
- W_6 \Tr{\chi^\dagger U + U^\dagger\chi} 
      \Tr{\rho^\dagger U + U^\dagger \rho}
- W'_6 \left[\Tr{\rho^\dagger U + U^\dagger \rho}\right]^2 \notag\\
& - L_7 \left[\Tr{\chi^\dagger U - U^\dagger\chi}\right]^2
- W_7 \Tr{\chi^\dagger U - U^\dagger\chi} 
      \Tr{\rho^\dagger U - U^\dagger \rho}
- W'_7 \left[\Tr{\rho^\dagger U - U^\dagger \rho}\right]^2 \notag\\
& - L_8 \Tr{\chi^\dagger U \chi^\dagger U + U^\dagger\chi U^\dagger\chi}
- W_8 \Tr{\rho^\dagger U\chi^\dagger U 
          + U^\dagger \rho \, U^\dagger\chi}\notag\\
& - W'_8 \Tr{\rho^\dagger U \rho^\dagger U 
            + U^\dagger \rho \, U^\dagger \rho}.
\end{align}

Next-to-leading order terms are given by tree-level contributions from
${\cal L}_4$ and one-loop contributions from ${\cal L}_2$. The latter
produce divergencies, which are treated by renormalizing the low energy
coefficients $L_i$, $W_i$ and $W'_i$.

First we consider $N_f=2$ flavours with non-degenerate quark masses.
Following \cite{FR-nondeg}, to ensure positivity of the fermion
determinant we implement the quark mass splitting and the chiral twist
with different generators $T_b$.  Because $T_3 = \tau_3 /2$ is usually
chosen for the twist, we use $T_1 = \tau_1 /2$ for the mass difference.
Defining
\begin{equation}
\chi_0 = B_0(m_d+m_u) \qquad \mathrm{and} \qquad \chi_1 = B_0(m_d-m_u) ,
\end{equation}
we have
\begin{equation}
\chi = \chi_0\mathds{1} + \chi_1 \tau_1
\end{equation}
and
\begin{equation}
\rho = \rho_0 \mathds{1} + \I\rho_3 \tau_3 \,,
\end{equation}
where $\rho_0=2W_0 a\cos\omega$ and $\rho_3=2W_0 a\sin\omega$. Due to
the properties of the Pauli matrices only the $L_7$ and $L_8$-terms
lead to contributions involving the mass splitting $\chi_1$.

For the minimum of the effective Lagrangian we obtain
\begin{equation}
\tilde\phi_1=\tilde\phi_2=0
\end{equation}
and
\begin{equation}
\tilde{\phi}_3 = F_0 \frac{\rho_3}{\chi_0} 
\left[ 1- \frac{8}{F_0^2} (2 L_{86} - W_{86}) \chi_0 \right] 
+ {\cal O}(a^2) \,,
\end{equation}
where terms of order $a^2$ can be neglected at this order of the
calculation, and we define
\begin{equation}
L_{ij}=L_i+N_fL_j \,.
\end{equation}

{}From a one-loop calculation of the renormalized propagator we obtain
the pion masses in next-to-leading order. Owing to the unequal quark
masses and the chiral twist the three masses are generally different.
For the component $\phi_2 \equiv \pi_2$ we get
\begin{align}
M_{\pi_2}^2 = m_\pi^2 & 
+ \frac{m_\pi^4}{32\pi^2F_0^2}\, 
  \ln\left(\frac{m_\pi^2}{\Lambda^2}\right)\notag\\
& + \frac{8}{F_0^2} \Big[- L^r_{54} \chi_0 (\chi_0 + \rho_0)
  - W^r_{54} ( \chi_0\rho_0 + \rho_0^2 + \rho_3^2 ) \notag\\
& + 2 L^r_{86} ( \chi_0^2 - \rho_3^2 ) 
  + 2 W^r_{86} ( \chi_0\rho_0 + \rho_3^2 ) 
  + 2 W'^r_{86} \, \rho_0^2 \Big],
\end{align}
where $m_\pi$ denotes the leading order pion mass,
\begin{equation}
m_\pi^2 = \chi_0 + \rho_0 + \frac{\rho_3^2}{2\chi_0} \,,
\end{equation}
$\Lambda$ is the renormalization scale, usually taken to be $4\pi F_0$,
and the superscript $r$ indicates the renormalized $L$, $W$ and $W'$
coefficients.

The masses of the other two pions can be read of from the mass
splittings
\begin{align}
M_{\pi_{1}}^2 - M_{\pi_2}^2 & 
= \frac{16}{F_0^2} L^r_{87} \chi_1^2 \,,\\
M_{\pi_{2}}^2 - M_{\pi_3}^2 & 
= \frac{16}{F_0^2} (L^r_{86}-W^r_{86}+W'^r_{86}) \rho_3^2 \,.
\end{align}
Setting $\chi_1=0$, this is consistent with the results for degenerate
quark masses \cite{MS,Sco}.  Moreover we observe
${\cal O}(a)$-improvement at maximum twist $\omega=\pi/2$, where 
$\rho_0=0$.

Note that with our choice of SU(2) generators, $\pi_1$ is the neutral
pion, and the charged pions are linear combinations of $\pi_2$ and
$\pi_3$. Due to the twist, however, the mass eigenstates are not the
usual charged pions but the components $\pi_2$ and $\pi_3$.

The pion decay constant $F_\pi$ is given by the matrix element
\begin{equation}
\langle 0 | A_{\mu,a} | \phi_b(\vec p)\rangle 
= \I F_\phi p_\mu \delta_{ab}\,,
\end{equation}
where $A_{\mu,a}$ is the axial current.  In next-to-leading order one
can determine the decay constant including order $a$ terms.  We get
\begin{equation}
\frac{F_\pi}{F_0} =
1 - \frac{1}{16\pi^2F_0^2} \,
m_\pi^2  \ln\left(\frac{m_\pi^2}{\Lambda^2}\right)
+ \frac{4}{F_0^2} (L^r_{54} \chi_0 + W^r_{54} \, \rho_0)
\end{equation}
without any contributions of the quark mass difference~$\chi_1$ and
coinciding with the result of \cite{MS}.

Now we turn to the consideration of $N_f=3$ flavours with degenerate
up- and down-quarks and a heavier strange-quark.  The generators of
SU(3) are given by the Gell-Mann matrices $T_b = \lambda_b /2$.
Defining $\chi_s=2 B_0 m_s$, the quark mass matrix is given by
\begin{equation}
\chi = \frac{1}{3}(2\chi_0+\chi_s)\mathds{1}
     - \frac{1}{\sqrt3}(\chi_s-\chi_0)\lambda_8 \,.
\end{equation}
For $N_f=3$ it is generally possible to introduce a twist with two
twist angles involving all flavours. In view of the fact that the
motivation for twist is to eliminate infrared problems stemming from
the light quark masses, and in order to ensure a positive quark
determinant \cite{Frezzotti2} we implement a chiral twist by a single
angle $\omega$ in the isospin sector of the lightest quarks, leading to
\begin{equation}
\rho 
= \left( \begin{array}{ccc}
\rho_0 + \I \rho_3 & 0 & 0 \\
0 & \rho_0 - \I \rho_3 & 0 \\
0 & 0 & |\rho|
\end{array} \right)
= \frac{1}{3} (2 \rho_0 + |\rho|)\mathds{1}
- \frac{1}{\sqrt{3}} (|\rho| - \rho_0) \lambda_8
+ \I \rho_3 \lambda_3
\end{equation}
with $|\rho| = 2 W_0\,a$.

The minimum of the potential is shifted again in $T_3$-direction to
\begin{equation}
\tilde{\pi}_3 =
F_0 \frac{\rho_3}{\chi_0} 
\left[ 1 - \frac{8}{F_0^2} \{ (2L_8-W_8) \chi_0 
+ (2L_6-W_6) (2\chi_0+\chi_s) \} \right] + {\cal O}(a^2) .
\end{equation}

At leading order the pseudoscalar meson masses are
\begin{alignat}{2}
m_\pi^2  & = \chi_0 + \rho_0 + \frac{\rho_3^2}{2\chi_0}\,, \\
m_K^2   & = \frac{1}{2} \left[(\chi_s+\chi_0)
 + (|\rho| + \rho_0) + \frac{\rho_3^2}{2\chi_0} \right], \\
m_\eta^2 & = \frac{1}{3} \left[(2\chi_s+\chi_0) + (2|\rho| + \rho_0)
 + \frac{\rho_3^2}{2\chi_0} \right].
\end{alignat}

Including the tree-level and one-loop terms at next-to-leading order
we obtain the pion masses
\begin{align}
M_{\pi_{1,2}}^2 = &\, m_\pi^2 + \frac{m_\pi^2}{96\pi^2 F_0^2}
 \left[ 3 m_\pi^2 \ln\left(\frac{m_\pi^2}{\Lambda^2}\right)
      - m_\eta^2 \ln\left(\frac{m_\eta^2}{\Lambda^2}\right)\right] 
  \notag\\
& + \frac{4}{F_0^2} \Big[ (4L^r_8+8L^r_6-2L^r_5-4L^r_4)\chi_0^2
+ (4L^r_6-2L^r_4)\chi_0\chi_s \notag\\
& + (-2L^r_5-4L^r_4+4W^r_8+8W^r_6-2W^r_5-4W^r_4)\chi_0\rho_0
+ (2W^r_6-2W^r_4)\chi_0|\rho| \notag\\
& + (-2L^r_4+2W^r_6)\chi_s\rho_0 \notag\\
& + (-2W^r_5-4W^r_4+4W'^r_8+8W'^r_6)\rho_0^2
+ (-2W^r_4+4W'^r_6)\rho_0|\rho| \notag\\
& + (-4L^r_8-8L^r_6+4W^r_8+8W^r_6-2W^r_5-4W^r_4)\rho_3^2
+ (-2L^r_6-L^r_4+2W^r_6)\rho_3^2\tfrac{\chi_s}{\chi_0} \Big] .
\end{align}
This expression shows that the presence of the strange quark,
whose mass term is not subject to any chiral twist here, foils the
${\cal O}(a)$-improvement at maximal twist, as expected.

The mass splitting, given by
\begin{equation}
M_{\pi_{1,2}}^2 - M_{\pi_3}^2 
= \frac{16}{F_0^2} (L^r_{86}-W^r_{86}+W'^r_{86})\rho_3^2 \,,
\end{equation}
is formally equivalent to that obtained for $N_f=2$, taking the
different Gasser-Leutwyler coefficients into account.

For the kaons we get
\begin{align}
M_K^2 =\,& m_K^2 + \frac{m_K^2}{48\pi^2F_0^2}\,
  m_\eta^2 \ln\left(\frac{m_\eta^2}{\Lambda^2}\right) \notag\\
& + \frac{2}{F_0^2} \Big[ (2L^r_8+8L^r_6-L^r_5-4L^r_4)\chi_0^2
  + (4L^r_8+16L^r_6-2L^r_5-6L^r_4)\chi_0\chi_s \notag\\
& + (2L^r_8+4L^r_6-L^r_5-2L^r_4)\chi_s^2 \notag\\
& + (-L^r_5-4L^r_4+2W^r_8+8W^r_6-W^r_5-4W^r_4)\chi_0\rho_0 \notag\\
& + (-L^r_5-4L^r_4+2W^r_8+6W^r_6-W^r_5-2W^r_4)\chi_0|\rho| \notag\\
& + (-L^r_5-2L^r_4+2W^r_8+6W^r_6-W^r_5-4W^r_4)\chi_s\rho_0 \notag\\
& + (-L^r_5-2L^r_4+2W^r_8+4W^r_6-W^r_5-2W^r_4)\chi_s|\rho| \notag\\
& + (-2W^r_5-6W^r_4+4W'^r_8+12W'^r_6)\rho_0 (\rho_0 + |\rho|) \notag\\
& + (-4L^r_8-8L^r_6+4W^r_8+8W^r_6-2W^r_5-6W^r_4+4W'^r_6)\rho_3^2 
  \notag\\
& + (-2L^r_8-6L^r_6+L^r_4+2W^r_8+6W^r_6-W^r_5-4W^r_4)
\rho_3^2\tfrac{\chi_s}{\chi_0}\Big],
\end{align}
and for the $\eta$-mass
\begin{align}
M_\eta^2 =\,& m_\eta^2 + \frac{m_\eta^2}{48\pi^2F_0^2}
 \left[ 3 m_K^2 \ln\left(\frac{m_K^2}{\Lambda^2}\right)
      - 2 m_\eta^2 \ln\left(\frac{m_\eta^2}{\Lambda^2}\right)\right] 
  \notag\\
& + \frac{m_\pi^2}{96\pi^2F_0^2}
 \left[ m_\eta^2 \ln\left(\frac{m_\eta^2}{\Lambda^2}\right)
      - 3 m_\pi^2 \ln\left(\frac{m_\pi^2}{\Lambda^2}\right)
      + 2 m_K^2 \ln\left(\frac{m_K^2}{\Lambda^2}\right)\right] \notag\\
& + \frac{4}{9F_0^2} \Big[
    (12L^r_8+24L^r_7+24L^r_6-2L^r_5-12L^r_4)\chi_0^2 \notag\\
& + (-48L^r_7+60L^r_6-8L^r_5-30L^r_4)\chi_0\chi_s
  + (24L^r_8+24L^r_7+24L^r_6-8L^r_5-12L^r_4)\chi_s^2 \notag\\
& + (-2L^r_5-12L^r_4+12W^r_8+24W^r_7+24W^r_6-2W^r_5-12W^r_4)
    \chi_0\rho_0 \notag\\
& + (-4L^r_5-24L^r_4-24W^r_7+30W^r_6-4W^r_5-6W^r_4)\chi_0|\rho| \notag\\
& + (-4L^r_5-6L^r_4-24W^r_7+30W^r_6-4W^r_5-24W^r_4)\chi_s\rho_0 \notag\\
& + (-8L^r_5-12L^r_4+24W^r_8+24W^r_7+24W^r_6-8W^r_5-12W^r_4)
    \chi_s|\rho| \notag\\
& + (-10W^r_5-24W^r_4+36W'^r_8+48W'^r_7+48W'^r_6)\rho_0^2 \notag\\
& + (-8W^r_5-30W^r_4-48W'^r_7+60W'^r_6)\rho_0|\rho|
  + (-24L^r_8-24L^r_7-24L^r_6 \notag\\
& + 24W^r_8+24W^r_7+24W^r_6-10W^r_5-24W^r_4+12W'^r_8+24W'^r_7+24W'^r_6)
    \rho_3^2 \notag\\
& + (24L^r_7-30L^r_6+9L^r_4-24W^r_7+30W^r_6-4W^r_5-24W^r_4)
\rho_3^2\tfrac{\chi_s}{\chi_0} \Big].
\end{align}

In addition to the masses we determine the decay constants for the
pseudoscalar meson octet.  The results are for the pions
\begin{align}
\frac{F_\pi}{F_0} = 1 & - \frac{1}{32\pi^2F_0^2} \left[
       2m_\pi^2\ln\left(\frac{m_\pi^2}{\Lambda^2}\right)
     +  m_K^2\ln\left(\frac{m_K^2}{\Lambda^2}\right) \right]\notag\\
  &  + \frac{4}{F_0^2} \left[ L^r_5\chi_0 + L^r_4(\chi_s+2\chi_0)
     + W_5^r\rho_0 + W_4^r(|\rho|+2\rho_0) \right],
\end{align}
for the kaons
\begin{align}
\frac{F_K}{F_0}  = 1 & - \frac{3}{128\pi^2F_0^2} \left[ 
       m_\pi^2\ln\left(\frac{m_\pi^2}{\Lambda^2}\right)
     + 2m_K^2\ln\left(\frac{m_K^2}{\Lambda^2}\right)
     +  m_\eta^2\ln\left(\frac{m_\eta^2}{\Lambda^2}\right) \right]
  \notag\\
  &  + \frac{2}{F_0^2} \left[ 
       L^r_5(\chi_0+\chi_s) + 2L^r_4(\chi_s+2\chi_0)
     + W_5^r(|\rho|+\rho_0) + 2W_4^r(|\rho|+2\rho_0) \right],
\end{align}
and for the $\eta$ meson
\begin{align}
\frac{F_\eta}{F_0} = 1 & - \frac{3}{32\pi^2F_0^2}\,
       m_K^2\ln\left(\frac{m_K^2}{\Lambda^2}\right) \notag\\
    & + \frac{4}{3F_0^2} \left[ L^r_5(2\chi_s+\chi_0) 
      + 3L^r_4(\chi_s+2\chi_0)
      + W_5^r(2|\rho|+\rho_0) + 3W_4^r(|\rho|+2\rho_0) \right].
\end{align}

In the limit of equal quark masses for all flavours, $\chi_s = \chi_0$,
and vanishing twist these expressions coincide to
\begin{equation}
\frac{F_\pi}{F_0} = 1 - \frac{N_f}{32\pi^2F_0^2} 
\left[ m_\pi^2\ln\left(\frac{m_\pi^2}{\Lambda^2}\right)\right]
+ \frac{4}{F_0^2} \left[ L^r_{54} \chi_0 + W^r_{54} \rho_0 \right] ,
\end{equation}
recovering the result of \cite{Rupak-Shoresh}.

As a check, in the continuum limit, $\rho=0$, our formulae for $N_f=3$
go over into those of \cite{GL2}.

The expressions given here can be used in the analysis of numerical
results from unquenched simulations of twisted mass lattice QCD with
two or three flavours. Fitting the data for masses and decay constants
with the chiral perturbation formulae will allow to extract numerical
values for the low energy coefficients \cite{DBW2,tmLQCD2}.

We thank I.~Montvay, F.~Farchioni and P.~Hofmann for discussions.

%
\end{document}